\documentstyle[twoside,psfig,amsmath]{article}
\begin{document}

\newcommand{\BS}{\mbox{\scriptsize BS}}

\setlength{\textheight}{7.7truein}

\markboth{\protect{\footnotesize\it M. Montero}}{\protect{\footnotesize\it PDE for a Stochastic Volatility Model}}

\thispagestyle{empty}
\setcounter{page}{1}

\vspace*{1in}

\centerline{\bf PARTIAL DERIVATIVE APPROACH FOR OPTION PRICING}
\baselineskip=13pt
\centerline{\bf IN A SIMPLE STOCHASTIC VOLATILITY MODEL}
\vspace*{0.37truein}

\centerline{\footnotesize MIQUEL MONTERO}
\baselineskip=12pt
\centerline{\footnotesize\it Departament de F\'{\i}sica Fonamental, Universitat de Barcelona,}
\baselineskip=10pt
\centerline{\footnotesize\it Diagonal 647, 08028 Barcelona, Spain}
\baselineskip=10pt
\centerline{\footnotesize\it E-mail: miquel\,.\,montero\,@\,ub\,.\,edu}

\vspace*{0.225truein}

\vspace*{0.21truein}
\begin{abstract}
We study a market model in which the volatility of the stock may jump at a random time $\tau$ from a fixed value $\sigma_a$ to another fixed value $\sigma_b$. This model was already described in the literature. We present a new approach to the problem, based on partial derivative equations, which gives a different perspective to the problem. Within our framework we can easily consider several prescriptions for the market price of volatility risk, and interpret their financial meaning. Thus, we recover solutions previously cited in the literature as well as obtain new ones.
\end{abstract}

\setcounter{footnote}{0}
\renewcommand{\thefootnote}{\alph{footnote}}

\vspace*{4pt}
\normalsize\baselineskip=13pt
\section{Introduction}
\noindent
The problem of pricing financial derivatives was already present in the aim of the early works in Mathematical Finance. Bachelier in 1900  proposed the arithmetic Brownian motion for the dynamical evolution of stock prices as a first step towards obtaining a price for options~\cite{C64}. Nevertheless the interest on this problem has increased remarkably in the past twenty years, after the publication of the works of Black and Scholes~\cite{BS73}, and Merton~\cite{M73}. The Black-Scholes model has been broadly used by practitioners thereafter, mainly due to its mathematical simplicity. It is well established, however, that this model fails to explain some statistical features shown in real markets. In particular, there are solid evidences pointing to the necessity of relaxing the assumption, present in the Black-Scholes model, that a constant volatility parameter drives the stock price. One of the tests more commonly used is based on a conceptually simple principle. Since the Black-Scholes price is a monotonous function on its arguments, the formula can be inverted in order to compute the {\it implied volatility\/}, the volatility that will reproduce the actual market conditions. The usual result is that the implied volatility is not constant, but a U-shaped function of the moneyness, whose minimum is at moneyness near to one ---{\it i.e.\/} when the current price of the underlying is close to the strike. This departure from the Black-Scholes model is known as the {\it smile effect\/}, and it is well documented in the literature~\cite{JR96}. 

Many models have been developed with the purpose of avoiding this inaccurate feature. We will mention here only a few of them. Merton itself~\cite{M76} proposed a model in which the volatility was a deterministic function of time. Cox and Ross~\cite{CR76} presented some alternative proposals that can be thought as models in which the volatility is stock-dependant. These and other similar contributions lead to a framework in which all the option risk comes from the fluctuations in the price of the underlying. In practical situations, however, it seems that this description is not sophisticated enough for explaining the actual changes in the level of volatility. Some authors have then suggested that the evolution of the volatility is driven by its own stochastic equation. Among these models of {\it stochastic volatility\/} we find works that are historically noteworthy: Hull and White~\cite{HW87} proposed a model where the squared volatility also follows a log-normal diffusion equation, independent of the stock price. Wiggins~\cite{W87} extended this idea and considered that the underlying and the volatility constitute a two-dimensional system of correlated log-normal random processes. Scott~\cite{S87}, but specially Stein and Stein~\cite{SS91} assumed that the instantaneous volatility follows a random mean-reverting process: an independent arithmetic Ornstein-Uhlenbeck process. Masoliver and Perell\'o~\cite{MP02} relaxed this assumption, and introduced correlation in the two-dimensional Wiener process. Heston~\cite{H93} turned the arithmetic model into a square-root correlated process.

All these seminal papers have in common that they model the stochastic behaviour of the volatility as a diffusion process. Naik~\cite{N93} developed a model in which the volatility can have only two known values, and the market switches back and forth between them, in a random way. This set-up can be used to model a market with high and low volatility periods.
Herzel~\cite{H98} studied a simplified version of this problem, in which the volatility, at the most, can jump once. This is a suitable model for encoding a market that may undergo a severe change in volatility only if some forthcoming event takes place. Since options have a limited lifetime, this seems not to be a very restrictive limitation. 
Herzel solved the problem of pricing the options using probability arguments, and showed that his model can account for the smile effect. 

We present here a different approach for obtaining fair option prices under Herzel's conditions. We will employ a technique of broad use both in research papers ({\it e.g.} in~\cite{H93}) and reference books ({\it e.g.} in~\cite{W98}) on this topic: we determine the partial derivative equations that the option price must fulfil, according to the It\^o convention, and solve it with the appropriate constrains. This scheme eases the task of considering several prescriptions for the market price of volatility risk, and leads to a plain way of interpreting the financial meaning of each of them.

The paper is structured as follows: in Sec.~2 we present the general market model and specify the differential equations that govern the traded securities. In Sec.~3 we study the way of obtaining a complete market. In Sec.~4 we explore the consequences of demanding that the market admits no arbitrage. In Sec.~5 we present explicit solutions for different market prices of the volatility risk. Section 6 contains actual numerical examples of these solutions and a financial interpretation of the results. The conclusions are drawn in Sec.~7. The paper ends with Appendix~A, where we detail the way we have followed for finding of one of the new solutions we have introduced.

\section{The Market Model}
\noindent
Let us begin with the general description of our set-up. We will assume that in our market there is at least a non-deterministic traded stock, $S$. The evolution of the price of this stock, from $S_0$ at $t=t_0$, is governed by the following differential equation\footnote{Throughout our exposition we will not specify the explicit dependence of the involved magnitudes, except if this may lead to confusion.} :   
\begin{equation*}
\frac{dS}{S}=\mu dt +\sigma dW,
\end{equation*}
where $W(t-t_0)$ is a one dimensional Brownian motion, with zero mean and variance equal to $t-t_0$, $\mu$ is a constant parameter, and $\sigma$, the volatility, is a stochastic quantity. The model assumes that the volatility have initially a given value $\sigma_a$, and that at most it may change to a different value $\sigma_b$ at instant $\tau>t_0$:  
\begin{equation}
\sigma(t;\tau)=\sigma_a {\mathbf 1}_{t< \tau} +\sigma_b {\mathbf 1}_{t\geq\tau}=
\sigma_a +(\sigma_b-\sigma_a) {\mathbf 1}_{t\geq\tau}, \label{sigma}
\end{equation}
where ${\mathbf 1}_{\{\cdot\}}$ denotes the indicator function, which assigns the value 1 to a true statement, and the value 0 to a false statement. The time $\tau$ in which such transition occurs is random and we will assume that it follows an exponential law:
\begin{equation*}
P(t_0<\tau\leq t)=1-e^{-\lambda (t-t_0)}. 
\end{equation*}
Note that with the previous definition, $\lambda$ is just the inverse of the mean transition time, $E[\tau]=\lambda^{-1}$.

We are also assuming that we will be capable of concluding whether the transition has taken place or not. This assumption does not imply that we can directly measure the value of $\sigma$, but that there exists a way to determine if $t \geq \tau$. This can be easily understood from the point of view of a practitioner. Let us suppose, for instance, that we are expecting that a relevant financial announcement is done. We do not know for sure when it will happen, but we belive that this new information will affect the level of volatility in our market. Even though we may not perform an instantaneous measure of the volatility in order to check the actual effect of the news, if they are published we will know it. We will return to this issue later.  

Upon the underlying stock $S$, we will define a new traded asset: the option $C$. The price of this option will depend explicitly on the moment $t_0$ in which we will decide to evaluate it, on the current stock price $S_0$ and on the level of volatility $\sigma_0$, but also on a set of peculiar parameters which we will label with a single symbol, $\kappa$. These parameters are the contract specifications that will characterize the option: the maturity time or the striking price, among other possibilities. This framework covers the European put and call options, e.g. the {\it vanilla\/} options or the binary options; also the American options, but does not include more exotic derivatives, such as the Asian options or the lookback options. 

The differential of the option price $C=C(t,S,\sigma;\kappa)$ has, according to the It\^o convention, the following expression:  
\begin{equation}
dC=\partial_t C dt+\partial_{S} C dS+\frac{1}{2} \sigma^2 S^2 \partial_{SS}^2 C dt+ \frac{\Delta C}{\sigma_b-\sigma_a} d \sigma, \label{dC}
\end{equation}
where
\begin{equation*}
\Delta C \equiv C(t,S,\sigma_b;\kappa)-C(t,S,\sigma_a;\kappa).
\end{equation*}
The last term in Eq.~(\ref{dC}) condenses the innovation with respect to the classical Black-Scholes expression, and represents the contribution of the randomness in the volatility to the dynamics of the option price. Note that this extra term is a product of the finite difference version of the derivative of $C$ with respect to $\sigma$, and $d\sigma$. In order to obtain an alternative expression for this object, we will simply differentiate Eq. (\ref{sigma}): 
\begin{equation}
d\sigma =(\sigma_b -\sigma_a) d{\mathbf 1}_{t\geq \tau}. \label{d_sigma}
\end{equation}
The differential of a indicator may seem a bizarre object. However, it is mathematically well defined, as we will shortly show. We can decompose this differential in two terms:
\begin{equation}
d{\mathbf 1}_{t\geq \tau}=\lambda  {\mathbf 1}_{t < \tau} dt - \lambda dG. \label{d1}
\end{equation}
The first term is regular, and the second involves function $G$,
\begin{equation*}
G=t{\mathbf 1}_{t < \tau} + \left(\tau-\frac{1}{\lambda}\right) {\mathbf 1}_{t\geq \tau},
\end{equation*}
which is proven to be a right continuous with left limits martingale. In order to ease the notation, nevertheless, we will keep the differential of the indicator in its early form, and use the referred decomposition only when it can clarify the problem.
We must stress however that 
$d{\mathbf 1}_{t\geq \tau}$ is a stochastic magnitude, independent of $dW$. Since $d\sigma$ does not directly contribute on the variation of the stock price $dS$, we can foreseen that there is a source of risk that cannot be explained in terms of the random evolution of the underlying asset. We postpone nonetheless the discussion of this issue, since it will be the matter of the next Section.  

Before that we want to point out that there is also a third kind of security traded in the market, a free-risk monetary asset $B$, which satisfies the corresponding equation: 
\begin{equation}
dB=r B dt. \label{dB}
\end{equation}
This security will allow us to borrow money when we need it, and it will provide a secure resort in order to keep the excess of cash if that is the case. In particular, it makes possible both the {\it self-financing strategy\/}, which allows closed portfolios, and the {\it net-zero investment\/}, the composition of a portfolio with no net value.

\section{Completeness of the Market}
\noindent
Let us face the problem of the completeness of the market. It is notorious that the market will be complete if we can construct for every security the so-called {\it replicating portfolio\/}, {\it i.e.\/}  a portfolio that mimics the behaviour of the asset. We have argued in the previous Section that not all the influence of $\sigma$ in the price of the option can be explained through $S$. We need then another security that can account for this component of the global risk. Instead of introducing a new traded asset depending only on $d\sigma$, with no clear financial interpretation, we have decided to use a {\it secondary\/} option $D(t,S,\sigma;\kappa')$: a derivative of the same nature of $C(t,S,\sigma;\kappa)$, but with a different set of contract specifications. This add-on completes the market if we are allowed to borrow money at a fixed interest rate whenever we need it, or to buy zero-coupon bonds in the case we obtain a surplus of cash. Thus we can write down $C$ as a combination of $\delta$ shares $S$, $\phi$ units of the riskless security $B$, and $\psi$ secondary options $D$:
\begin{equation*}
C=\delta S + \phi B + \psi D.
\end{equation*}
The variation in the value of both portfolios fulfills
\begin{equation*}
dC=\delta dS + \phi dB + \psi dD,
\end{equation*}
where we have taken into account two capital facts. On one hand $\delta$, $\phi$ and $\psi$ are nonanticipating functions of $S$ and $D$, e.g. $d\delta$, $d\phi$ and $d\psi$ do not depend on the new random information in $dW$ and $d\sigma$. On the other hand, we adopt a {\it self-financing strategy\/}, in which there is no net cash flow entering or leaving the replicating portfolio~\cite{HP81}: 
\begin{equation*}
S d\delta  +  B d\phi+ D d\psi =0.
\end{equation*}

We will replace $dC$ by the expression in~(\ref{dC}), and we will take into account the properties shown in~(\ref{d_sigma}) and~(\ref{dB}), to finally obtain: 
\begin{equation}
\partial_t C dt+\partial_{S} C dS+\frac{1}{2} \sigma^2 S^2 \partial_{SS}^2 C dt+ \Delta C d{\mathbf 1}_{t\geq \tau}=\delta dS + r \phi B dt + \psi dD. \label{PDE_1}
\end{equation}
We can proceed with $dD$ in an analogous way,
\begin{equation}
dD=\partial_t D dt+\partial_{S} D dS+\frac{1}{2} \sigma^2 S^2 \partial_{SS}^2 D dt+ \Delta D d{\mathbf 1}_{t\geq \tau}, \label{dD}
\end{equation}
where the natural definition of $\Delta D$,
\begin{equation*}
\Delta D=D(t,S,\sigma_b;\kappa')-D(t,S,\sigma_a;\kappa'),
\end{equation*}
has been used. In order to recover a deterministic partial differential equation we must guarantee that all the terms containing the 
stochastic magnitudes $dS$ and $d{\mathbf 1}_{t\geq \tau}$ mutually cancel out. Thus we must demand that
\begin{equation*}
\partial_{S} C = \delta + \psi \partial_{S} D,
\end{equation*}
condition named {\it delta hedging\/}, and also that
\begin{equation*}
\Delta C =  \psi\Delta D,
\end{equation*}
which is usually referred as {\it vega hedging\/}, or sometimes as {\it psi hedging\/}~\cite{MP03}. 

The previous hedging conditions reduce Eq.~(\ref{PDE_1}) to
\begin{equation}
\partial_t C dt+\frac{1}{2} \sigma^2 S^2 \partial_{SS}^2 C dt = r \phi B dt + \frac{\Delta C}{\Delta D} \left( \partial_t D dt+\frac{1}{2} \sigma^2 S^2 \partial_{SS}^2 D dt\right), \label{CBnD}
\end{equation}
expression that still involves $B$, which is not an inner variable of  the option prices $C$ and $D$. This problem can be fixed using together the definition of the portfolio and the {\it psi hedging\/},
\begin{equation*}
\phi B=C-\delta S - \psi D=C -\left(\partial_{S} C - \frac{\Delta C}{\Delta D}\partial_{S} D \right)S - \frac{\Delta C}{\Delta D} D.
\end{equation*}
Thus, the replacement of $\phi B$ in Eq.~(\ref{CBnD}) leads to
\begin{equation*}
\partial_t C +\frac{1}{2} \sigma^2 S^2 \partial_{SS}^2 C -rC +rS\partial_S C
=\frac{\Delta C}{\Delta D} \left( \partial_t D+\frac{1}{2} \sigma^2 S^2 \partial_{SS}^2 D-rD +rS\partial_S D\right).
\end{equation*}
This formula implies the existence of an arbitrary function $\chi=\chi(t,S,\sigma)$, which uncouples the problem of finding $C$ and $D$:
\begin{equation}
\chi(t,S,\sigma)=\frac{1}{\Delta C}\left(
\partial_t C +\frac{1}{2} \sigma^2 S^2 \partial_{SS}^2 C -rC +rS\partial_S C \right). \label{chi}
\end{equation}
Obviously the same formula, just replacing $C$ by $D$, is valid for the secondary option. This fact proves that the option $D$ completes the market indeed~\cite{BR96,MP03}. Note that if $\chi(t,S,\sigma)=0$ we recover the classical Black-Scholes equation.

\section{Arbitrage-free scenario}
\noindent
We have to state some criterion before we can choose a valid candidate for $\chi(t,S,\sigma)$. We will first determine the meaning of this arbitrary function. Consider $Y$, a portfolio which involves shares, bonds, one primary option, and secondary options:   
\begin{equation*}
Y=C+\bar{\delta} S + \bar{\phi} B + \bar{\psi} D.
\end{equation*}
Moreover, the relative amount of each security is such that, at the beginning, the portfolio has no net value, {\it i.e.\/}  $Y=0$. In fact, we will also demand that the changes in the value of the portfolio are not coming from a cash flow. The nonanticipating nature of $\bar{\delta}$, $\bar{\phi}$, and $\bar{\psi}$ makes that $dY$ takes the following form:   
\begin{equation*}
dY=dC+\bar{\delta} dS - r\left(C + \bar{\delta}S +\bar{\psi} D\right) dt. 
\end{equation*}
Finally, we can use Eqs.~(\ref{dC}),~(\ref{d_sigma}),~(\ref{dD}), and~(\ref{chi}) in order to obtain:
\begin{equation}
dY=(\Delta C+\bar{\psi} \Delta D)\left(\chi dt + d{\mathbf 1}_{t\geq \tau}\right) \label{dY} 
\end{equation}
where we have removed all the dependence in $dS$, just setting $\bar{\delta}= -\partial_S C-\bar{\psi} \partial_S C$.
In this case, we must ensure that either $dY=0$, or $dY$ has no definite sign. We will avoid the  choice $\bar{\psi}=-\Delta C/\Delta D$, which leads back to the trivial case $dY=0$. If $dY\geq0$, and $dY\neq 0$, we will have to design a null portfolio whose value can only rise. In other words, this market shows arbitrage opportunities. Obviously, the reciprocal scenario, {\it i.e.\/}  $dY\leq0$ and $dY\neq 0$, also lead to arbitrage, just building the $\bar{Y}=-Y$ portfolio. Clearly, the arbitrage possibilities other portfolios can be easily translated into $Y$ terminology. Therefore we need to analyse the behaviour of $dY$, and thus we will find the constrains to $\chi(t,S,\sigma)$.  Now it will be very convenient to recall the decomposition of $d{\mathbf 1}_{t\geq \tau}$ stated in Eq.~(\ref{d1}),
\begin{equation}
dY=(\Delta C+\bar{\psi} \Delta D)\left(\left(\chi + \lambda  {\mathbf 1}_{t < \tau}\right) dt - \lambda dG\right), \label{dYnG} 
\end{equation}
and to inspect the properties of $dG$: 
\begin{equation*}
d G = \left\{
\begin{array}{lr} 
0 & t \geq \tau, \\ 
dt -\lambda^{-1}&t<\tau \leq t+dt, \\
dt& \tau > t+dt.\\
\end{array}\right.
\end{equation*}
It is clear that for $t\geq\tau$, $dY$ reduces to $dY=(\Delta C+\bar{\psi} \Delta D) \chi dt$. But once the jump has happened, there is no financial reason to have a price that differs from the Black-Scholes price corresponding to $\sigma=\sigma_b$. And this is what we will get if we set $\chi =0$ right after the change in the volatility.

When the jump has not yet happened the differential $dY$ reads: 
\begin{equation*}
dY=(\Delta C+\bar{\psi} \Delta D)\left(\chi dt + \lambda (dt - dG)\right),  
\end{equation*}
with $\left(dt - dG\right)\geq0$. Therefore we must choose $\chi <0$ for $t< \tau$. Collecting all this we obtain the following formula:
\begin{equation*}
\chi(t,S,\sigma)=-\Omega(t,S)\, {\mathbf 1}_{\sigma=\sigma_a}=-\Omega(t,S)\, {\mathbf 1}_{t < \tau},
\end{equation*}
where $\Omega(t,S)$ is a strictly positive-definite bounded function depending on $t$ and $S$. This function $\Omega$ may depend, in a parametric way, on $\sigma_a$ and $\sigma_b$, but as well on $t_0$ and $S_0$. In fact, in general it may also depend on some parameters among those that characterise the contract specifications but, and this is a crucial point, never on all of them. We must have in mind that Eq.~(\ref{chi}) must hold at least for another option $D$, different from $C$. Otherwise the market will not be complete.  

We have shown the mathematical properties that $\chi(t,S,\sigma)$ must fulfil although we have deepened very little in its financial interpretation. Let us introduce function $\Psi(t,S,\sigma)$,
\begin{equation*}
\Psi(t,S,\sigma)=\left(\lambda-\Omega(t,S)\right)\, {\mathbf 1}_{t < \tau},
\end{equation*}
in Eq.~(\ref{dYnG}), 
\begin{equation*}
dY=(\Delta C+\bar{\psi} \Delta D)\left(\Psi dt - \lambda dG\right),  
\end{equation*}
and then evaluate the conditional expectation of $dY$, for a given value of $S$. Since $E[dG]=0$, it is clear that $E[dY|S]= E[\Psi|S] (\Delta C+\bar{\psi} \Delta D) dt$. Thus $\Psi(t,S,\sigma)$ measures the market price of the volatility risk, and it is exogenous to the market itself. It should be the financial agents who determine this function on the basis of their own appreciation of the actual risk. 
For instance, some authors~\cite{MP03} demand the absence of the so-call {\it statistical arbitrage\/}, that is $E[dY|S]=0$. This requirement implies that $\Psi=0$, {\it i.e.\/}  that $\Omega=\lambda$.

\section{Explicit solutions}
\noindent
We can now solve Eq.~(\ref{chi}) under appropriate conditions. For example, we shall begin assuming that $\Omega=\bar{\lambda}$ is constant, but not necessarily equal to $\lambda$, 
\begin{equation}
\partial_t C +\frac{1}{2} \sigma^2 S^2 \partial_{SS}^2 C -rC +rS\partial_S C +\bar{\lambda} \Delta C  {\mathbf 1}_{t < \tau}=0. \label{lambda}
\end{equation}
This implies that the risk is felt uniform in time. We will also consider that the price of the option is constrained by the final condition:
\begin{equation*}
C(T,S,\sigma;K)=\Phi(S;K),
\end{equation*}
which means that it will be a European-style option, where the price of the derivative in a fixed instant in the future, the maturity time, only depends on the actual value of the underlying at that moment and on some reference value, the strike, $K$. The function $\Phi$, the payoff, will change for different kind of options within this same family. For instance, for the plain {\it vanilla\/} call we have:
\begin{equation*}
C(T,S,\sigma;K)=\max(S(T)-K,0).
\end{equation*}
In addition, the mathematical nature of Eq.~(\ref{chi}) demands that the solution satisfy two extra boundary conditions which, in this case, read 
\begin{equation*}
C(t,0,\sigma;K)=0, \mbox{ and,} \lim_{S\rightarrow \infty} \frac{C(t,S,\sigma;K)}{S}=1.
\end{equation*}
Obviously, the same procedure can be used for other $\Phi$ functions, such as the binary call where the payoff is
\begin{equation*}
C(T,S,\sigma;K)={\mathbf 1}_{S(T) \geq K},
\end{equation*}
where the boundary conditions to be fulfilled are
\begin{equation*}
C(t,0,\sigma;K)=0, \mbox{ and,} \lim_{S\rightarrow \infty} C(t,S,\sigma;K)=1.
\end{equation*}
Therefore, we will not specify a single function $\Phi$, but we will treat all the suitable candidates at once. Moreover, we will use the term ``Black-Scholes price", $C^{\BS}$, as a synonymous of the solution of the Black-Scholes equation for the given payoff, without further distinction.

This will be the case when considering Eq.~(\ref{lambda}) for $\tau \leq t$, since then it reduces to the Black-Scholes model:
\begin{equation*}
\partial_t C +\frac{1}{2} \sigma^2_b S^2 \partial_{SS}^2 C -rC +rS\partial_S C=0,
\end{equation*}
whose solution is accordingly $C(t,S,\sigma_b;K)=C^{\BS}(t,S,\sigma_b;K)$. Nevertheless, we will show the main guidelines to solve it, because
this will illustrate more sophisticated problems to come. The first step is to introduce two new variables, $t^*=T-t$ and $x=\log(S)+\left(r-\sigma_b^2/2\right)(T-t)$, and to assume that $C$ depends on its own arguments only through them: 
\begin{equation*}
C(t,S,\sigma_b;K)=e^{-r(T-t)} V\left(T-t,\log(S)+\left(r-\frac{\sigma_b^2}{2}\right)(T-t);K\right).
\end{equation*}
This assumption implies the existence of a function of two variables $V(t^*,x;K)$ that obeys the following differential equation:
\begin{equation}
\partial_{t^*}  V=\frac{1}{2} \sigma_{b}^2 \partial_{xx}^2 V. \label{V} 
\end{equation}
Note that $t^*$ represents a reversion of the time arrow, that starts now at maturity. We have thus transformed our final condition into an initial one: 
\begin{equation*}
V(0,x_0;K)=\Phi(e^{x_0};K).
\end{equation*}
This problem has a straightforward solution:
\begin{equation}
V(t^*,x;K)=\int_{-\infty}^{+\infty} dx_0 \Phi(e^{x_0};K)\frac{1}{\sqrt{2\pi \sigma_{b}^2 t^*}} e^{-\frac{(x-x_0)^2}{2 \sigma_{b}^2 t^*}}, \label{Vsol}
\end{equation}
and therefore,
\begin{equation}
C(t,S,\sigma_b;K)=e^{-r(T-t)} \int_{-\infty}^{+\infty} dx_0 \Phi(e^{x_0};K)\frac{1}{\sqrt{2\pi \sigma_{b}^2 (T-t)}} e^{-\frac{\left(\log(S)+\left(r-\sigma_b^2/2\right)(T-t)-x_0\right)^2}{2 \sigma_{b}^2 (T-t)}}, \label{CbH}
\end{equation}
which is the Black-Scholes price. When $\tau > t$ the equation for $C(t,S,\sigma_a;K)$ is a little more complex:
\begin{equation*}
\partial_t C +\frac{1}{2} \sigma^2 S^2 \partial_{SS}^2 C -rC +rS\partial_S C +\bar{\lambda} \left(C^{\BS}(t,S,\sigma_b;K)-C\right)=0.
\end{equation*}
The last term comes from the $\chi \Delta C$ contribution. The key point is to realize that in the expression for $\Delta C$ appears, not only $C(t,S,\sigma_b;K)$, which we have found in Eq.~(\ref{CbH}), but also $C(t,S,\sigma_a;K)$, the unknown quantity. The procedure to follow is very similar to the one of the previous case. We will use again variable $t^*$, and define $\xi$ as $\xi=\log(S)+\left(r-\sigma_a^2/2\right)(T-t)$. In fact $x$ relates to $\xi$ through $x=\xi+\left(\sigma_a^2-\sigma_b^2\right)t^*/2$, what will be useful in a forthcoming step. Now we assume again a particular dependence on the price of this new variables, 
\begin{equation*}
C(t,S,\sigma_a;K)=e^{-(r+\bar{\lambda})(T-t)} Z\left(\log(S)+\left(r-\frac{\sigma_a^2}{2}\right)(T-t),T-t\right),
\end{equation*}
where $Z(t^*,\xi;K)$ obeys the following equation,
\begin{equation*}
\partial_{t^*}  Z=\frac{1}{2} \sigma_{a}^2 \partial_{\xi\xi}^2 Z +\bar{\lambda} e^{\bar{\lambda} t^*} V\left(\xi+\frac{\sigma_a^2-\sigma_b^2}{2}t^*,t^*;K\right), 
\end{equation*}
with the function $V$ of Eq.~(\ref{Vsol}); and the corresponding initial condition,
\begin{equation*}
Z(0,\xi_0;K)=\Phi(e^{\xi_0};K).
\end{equation*}
After some algebra its solution reads
\begin{equation*}
C(t,S,\sigma_a;K)=e^{-\bar{\lambda}(T-t)} C^{\BS}(t,S,\sigma_a;K) +\bar{\lambda} \int_{t}^{T} du\, e^{-\bar{\lambda}(u-t)} C^{\BS}(t,S,\bar{\sigma}(u-t,T-t);K),
\end{equation*}
where some short of ``effective variance", $\bar{\sigma}(t_a,t_b)$, has been introduced:
\begin{equation}
\bar{\sigma}^2(t_a,t_b)\equiv \frac{\sigma_a^2 t_a +\sigma_b^2(t_b-t_a)}{t_b}.\label{bar_s}
\end{equation}
Note that $\bar{\sigma}(0,T-t)=\sigma_b$ and $\bar{\sigma}(T-t,T-t)=\sigma_a$.
This behaviour can be used for compacting the solution. We can perform a typical integration by parts inside the integral sign and recover:
\begin{equation}
C(t,S,\sigma_a;K)=C^{\BS}(t,S,\sigma_b;K) + \int_{t}^{T} du\, e^{-\bar{\lambda}(u-t)} \partial_u C^{\BS}(t,S,\bar{\sigma}(u-t,T-t);K).\label{CaH}
\end{equation}
The main benefit of the last expression is that it can be easily combined with Eq.~(\ref{CbH}), thus yielding:
\begin{equation}
C(t,S,\sigma;K)=C^{\BS}(t,S,\sigma_b;K) +{\mathbf 1}_{t < \tau} \int_{t}^{T} du\, e^{-\bar{\lambda}(u-t)} \partial_u C^{\BS}(t,S,\bar{\sigma}(u-t,T-t);K).\label{CH}
\end{equation}
Note that this result is not sensitive to whether $\tau$ is smaller than $T$, or not. Depending on the payoff function, the integral that appears in Eq.~(\ref{CH}) can be computed, and analytic expressions for the option price can be obtained. In the most of the cases we have explored, however, the final formulas are cumbersome, thus providing little insight into the problem. We will present the simplest expression we have found by way of example. It corresponds to a {\it vanilla\/} call, in the special case that the {\it discounted moneyness\/} is equal to one, {\it i.e.\/}  $S=K e^{-r(T-t)}$, and that $\bar{\lambda}$, $\sigma_a$ and $\sigma_b$ are such that they fulfil
\begin{equation*}
\bar{\lambda}=\frac{\sigma^2_b-\sigma^2_a}{8}>0.
\end{equation*}
Then Eq.~(\ref{CH}) reduces to
\begin{eqnarray*}
C(t,K  e^{-r(T-t)},\sigma;K)&=&C^{\BS}(t,K e^{-r(T-t)},\sigma_b;K) \\
&-&{\mathbf 1}_{t < \tau}\left(\sigma_b-\sigma_a\right) K e^{-r(T-t)+\sigma^2_b (T-t)/8}  \sqrt{\frac{T-t}{2 \pi}}. 
\end{eqnarray*}

Up to this point we have reproduced the framework that corresponds to the problem stated by Herzel~\cite{H98}. Our output agrees with his expression for $t=0$, which is in fact the first option price given in Herzel's paper, although he does not number the equation. After that, he generalizes his formulation for any later instant of time, $0\leq s\leq T$. Unfortunately, there is an erratum in his proposal. Thus, the limits in the definite integral in Eq.~(4.28) should be $0$ and $T-s$, instead of $s$ and $T$. Or, in an equivalent way, $t$ should be replaced by $t-s$, keeping the rest unchanged, including $dt$.

More general solutions can be obtained using the same approach, with little extra effort. We can consider, for instance, the case of a $\chi$ depending on all the involved time magnitudes:
\begin{equation*}
\chi=-\eta(t;t_0,T) {\mathbf 1}_{t < \tau},
\end{equation*}
with $\eta(t;t_0,T)>0$. The solution for $t=t_0$ is simply\footnote{Given that, up to this moment, the notation did not induce to misunderstanding, we had not stressed the difference between $t_0$, the actual time in which the options is evaluated, and $t$, a generic instant of time, $t_0\leq t \leq T$.} :
\begin{eqnarray}
C(t_0,S_0,\sigma;K)&=&C^{\BS}(t_0,S_0,\sigma_b;K) \nonumber \\
&+&{\mathbf 1}_{t_0 < \tau} \int_{t_0}^{T} du\, e^{-\int_{t_0}^{u}dt'\eta(t';t_0,T)} \partial_u C^{\BS}(t_0,S_0,\bar{\sigma}(u-t_0,T-t_0);K). \nonumber \\ \label{Ceta}
\end{eqnarray}
We have thus obtain a broad set of valid prices with no clear financial interpretation. A plausible requirement that may help us to discard candidates is to demand that the final solution only depends on $T-t_0$. 
This forces that $\eta(t;t_0,T)=f(t-t_0;T-t_0)$. In Appendix A we find that, if we follow an heuristic approach that tries (in vain) to cancel out all the risk, the option price is
\begin{eqnarray}
C(t_0,S_0,\sigma;K)&=&C^{\BS}(t_0,S_0,\sigma_b;K)\nonumber \\ 
&+& {\mathbf 1}_{t < \tau} \int_{t_0}^{T} du\, \frac{1+e^{-\lambda(u-t_0)}}{2} \partial_u C^{\BS}(t_0,S_0,\bar{\sigma}(u-t_0,T-t_0);K). \nonumber \\ \label{CM}
\end{eqnarray}
We will disregard at this moment the possible interpretation of this solution, and concentrate in its validity instead. It is straightforward to check that if we replace
\begin{equation}
\eta(t;t_0,T)=\eta(t-t_0)=\lambda \frac{e^{-\lambda(t-t_0)}}{1+e^{-\lambda(t-t_0)}}, \label{H_eta}
\end{equation}
in Eq.~(\ref{Ceta}), we will recover Eq.~(\ref{CM}).

\section{Numerical computation}
\noindent
Now it is time to analyse and compare the different solutions we have found, and eventually to represent some of them. This task is easier if we express our results in the form of the expected value of the discounted payoff, under some appropriate probability density function:
\begin{equation}
C(t_0,S_0,\sigma_0;K)=E^Q[e^{-r(T-t_0)} \Phi(S(T);K)]. \label{EQ}
\end{equation}
Thus we will be able to translate the functional form of $\chi$ into an equivalent model. Since all the possible candidates to be the fair price collapse to the Black-Scholes solution if the jump takes place, we will assume that $t_0<\tau$ from now on.

\begin{figure}[htbp] 
\vspace*{13pt}
\centerline{\psfig{file=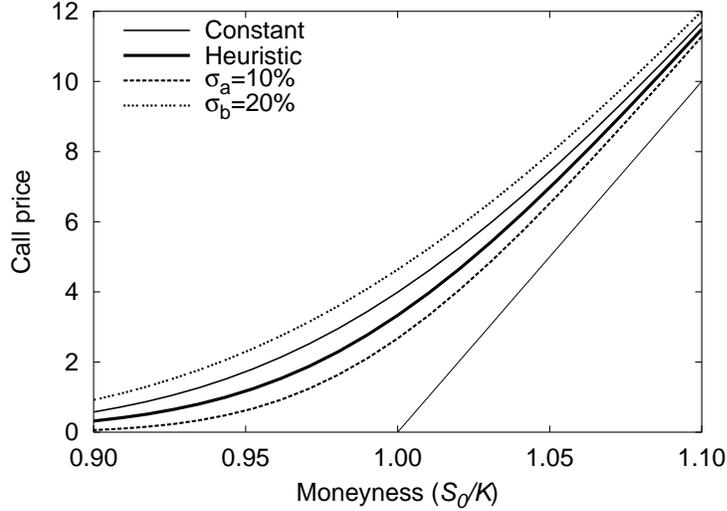,height=7truecm}}
(a)
\centerline{\psfig{file=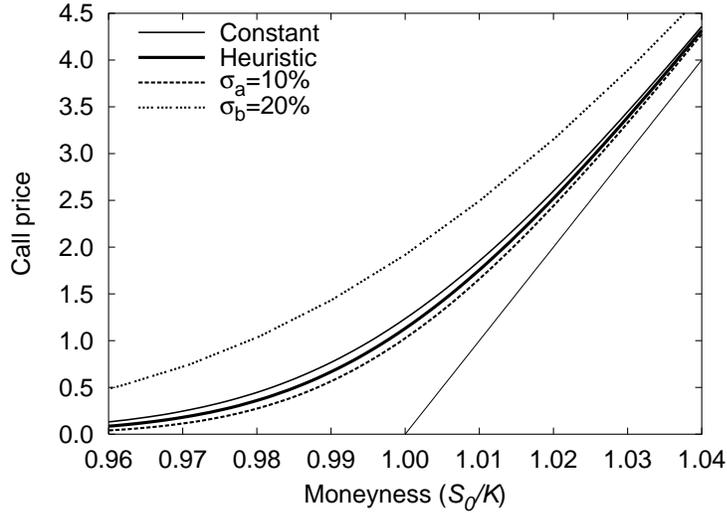,height=7truecm}}
(b)
\vspace*{13pt}
\caption{Option pricing in terms of the present moneyness, $S_0/K$ for two different maturities: (a) $T-t_0=0.25$ years, and (b) $T-t_0=0.05$ years. The depicted results correspond to a vanilla call, $\Phi(S;K)=\max(S-K,0)$. The numerical value of the involved parameters are $r=5\%$, $\sigma_a=10\%$, $\sigma_b=20\%$, $\lambda^{-1}=0.1$ years, and $K=100$, in suitable currency units. The ``constant" line correspond to setting $\eta=\lambda$, whereas by ``heuristic" we mean the choice in Eq.~(\ref{H_eta}), which was somewhat inspired by an heuristic approach. Note that the first price is more similar to the plain Black-Scholes price with $\sigma=\sigma_b$, and that conversely the second method leads to a price closer to the Black-Scholes one for $\sigma=\sigma_a$. The discrepancy is reduced as the maturity time approaches. All the plots were obtained using Monte Carlo techniques over $100\,000$ replicas.}
\label{fig1}
\end{figure}

Let us begin with $\eta=\bar{\lambda}$. In that case the final price $S(T)$ in Eq.~(\ref{EQ}) can be expressed in the following terms
\begin{eqnarray*}
S(T)&=&S_0 e^{\left(r-\sigma_a^2/2\right)(\bar{\tau}-t_0)+\sigma_a \overline{W}(\bar{\tau}-t_0)} {\mathbf 1}_{t < \bar{\tau}}\\
&+& S(\bar{\tau}) e^{\left(r-\sigma_b^2/2\right)(T-\bar{\tau})+\sigma_b \left[\overline{W}(T-t_0)-\overline{W}(\bar{\tau}-t_0)\right]}{\mathbf 1}_{t \geq \bar{\tau}},
\end{eqnarray*}
where, as usual, $\mu$ has been replaced by $r$, and $\overline{W}(t-t_0)$, a new 
Brownian motion with zero mean and variance equal to $t-t_0$, has been introduced. The jump process $\bar{\tau}$ follows also a different exponential law: 
\begin{equation*}
P(t_0<\bar{\tau}\leq t)=1-e^{-\bar{\lambda} (t-t_0)}. 
\end{equation*}
Thus when $\bar{\lambda}\neq \lambda$ we will consider in practice that the model is not accurate in the forecast of the actual mean transition time, and that it should be replaced by another value. It is not our intention to add superfluous complexity to the study of the several proposals we have done. Therefore we will concentrate in the case that the original $\lambda$ (and $\tau$) is used.

Summing up, we can actually compute the price of the option under the previous assumptions with the following procedure. We choose a value for $\tau$ that follows the proper probability density. If that time is bigger than the maturity we need to generate a zero mean Gaussian variable, with $T-t_0$ variance. Thus $S(T)$ will be governed only by $\sigma_a$. Conversely, if $\tau<T$ we will need two independent zero mean Gaussian variables, with variances equals to $\tau-t_0$ and $T-\tau$. In this case $\sigma_a$ governs the behaviour of the equivalent stock price up to $S(\tau)$, and $\sigma_b$ does thereafter. The averaged value of the discounted payoffs will lead to the correct estimation of the desired magnitude. This set-up is the appropriate to perform Monte Carlo numerical simulation, as it is shown in Fig.~\ref{fig1}.

\begin{figure}[htbp]
\vspace*{13pt}
\centerline{\psfig{file=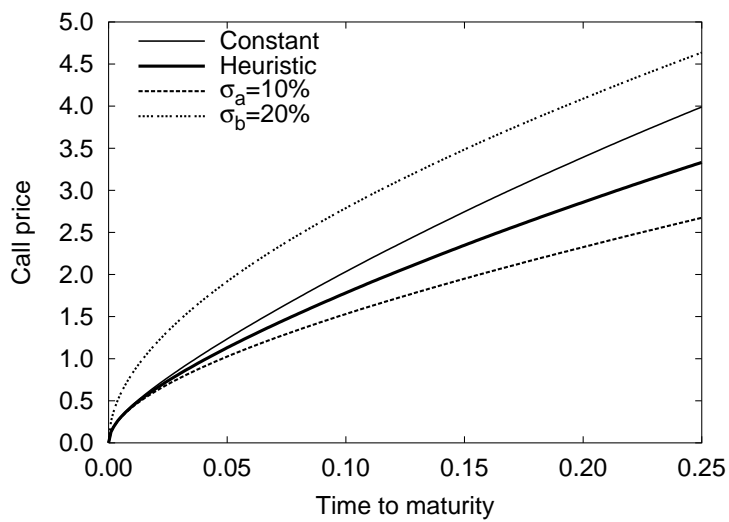,height=7truecm}}
\vspace*{13pt}
\caption{Option pricing in terms of time to maturity when the present moneyness is one. We consider an at-the-money call in the same set-up described in Fig.~\ref{fig1}. The plot shows how both prices change when the time to expiration shortens. Clearly the price that fully ignores the risk of a change in the volatility becomes more accurate as the probability of a jump reduces.}
\label{fig2}
\end{figure}

The choice in Eq.~(\ref{H_eta}) for $\eta$ is equivalent in law to consider that
\begin{eqnarray*}
S(T)&=&S_0 e^{\left(r-\sigma_a^2/2\right)(\tau-t_0)+\sigma_a \overline{W}(\tau-t_0)} {\mathbf 1}_{t < \tau}\\
&+&  \frac{1}{2}S(\tau)e^{\left(r-\sigma_a^2/2\right)(T-\tau)+\sigma_a \left[\overline{W}(T-t_0)-\overline{W}(\tau-t_0)\right]}{\mathbf 1}_{t \geq \tau}\\
&+& \frac{1}{2}S(\tau)e^{\left(r-\sigma_b^2/2\right)(T-\tau)+\sigma_b \left[\overline{W}(T-t_0)-\overline{W}(\tau-t_0)\right]}{\mathbf 1}_{t \geq \tau},
\end{eqnarray*}
with the same specifications for $\tau$ and $\overline{W}(t-t_0)$ as in the past scenario. Note that this particular choice for the market assessment of risk leads to the following alternative strategy. We choose again a value for $\tau$, using an exponential probability density function of mean $1/\lambda$. If that time is bigger than the maturity, nothing changes with respect to the previous example, and $\sigma_a$ drives the evolution of the underlying all the time. But when $\tau<T$ the picture changes. The price $S(T)$ is the arithmetic mean of the two possible paths: one in which the volatility is $\sigma_b$ right after the jump, and the other that considers that $\sigma_a$ remains unchanged. This is the reason to dissociate the premise that assumes the existence of a distinctive time value $\tau$, and the innovation that it carries. We can be sure that forthcoming news may affect the market and, at the same time, only guess about the final effect. Thus Eq.~(\ref{H_eta}) leads to a more conservative risk analysis, in the sense that this price is near the Black-Scholes value corresponding to $\sigma_a$, whereas the $\eta=\lambda$ choice anticipates more intensely the future change in the volatility. This explains the behaviour of the different call prices observed in Fig.~\ref{fig1}. Also in this figure, but specially in Fig.~\ref{fig2}, we can check that the two prices converge to the {\it no-jump\/} solution as the maturity horizon comes closer.

Finally, we want to point out that we can obtain solutions in all the range of confidence levels for the model. Thus, if we set $q$ equals to the probability that the volatility actually varies when the jump takes place, the function $\eta$ that express this risk evaluation is 
\begin{equation*}
\eta(t;t_0,T)=\eta(t-t_0)=\lambda \frac{e^{-\lambda(t-t_0)}}{(1-q)/q+e^{-\lambda(t-t_0)}}.
\end{equation*}
The only constrain is that we cannot neglect all the risk just setting $q=0$. Therefore $0<q\leq1$ leads to a valid price, although in absence of further information $q=1/2$ and $q=1$ seems to be the only privileged values.

\section{Conclusions}
\noindent
We have revisited the framework stated by Herzel, in which the dynamics of one asset $S$ is driven by a log-normal diffusion equation with an stochastic volatility parameter $\sigma$. The volatility of this stock may jump at a random time $\tau$ from a fixed initial value $\sigma_a$ to another fixed final value $\sigma_b$. And no more than one of such jumps is allowed. This event can model, for instance, the future publication of crucial information related to this specific market. 

We have introduced a procedure for obtaining fair option prices, different from the used one in Herzel's manuscript. There, the author exploits intensively probability arguments for finding the necessary and sufficient conditions that the model must fulfil to be complete and arbitrage-free. Thus he derive the equivalent martingale measure $Q$. We have employed another technique of broad use in this field. We have determine the partial derivative equations that, according to the It\^o convention, the option price must fulfil. We have shown that the use of a {\it secondary option\/} completes the market. After that, we have demanded that the market has no arbitrage and we have found the exogenous function that 
measures the market price of the volatility risk. We have explored the output for several choices of this function and, incidentally, we have amended some of the results presented in the original reference, where the risk premium was null.

In fact, one of the biggest benefits of our approach is when considering more sophisticated prescriptions for the market price of volatility risk. We have not only obtained closed formulas in such a cases, but we have also been able to interpret the financial meaning of them. We have seen how a choice for the volatility risk price can be translated into a lack of confidence in the model premises. For instance, a constant risk price, other than zero, plays the same role of a redefinition in the mean transition time of the jump process.

In particular, we have studied with some detail a solution that can be understood as the response of a suspicious market maker, who admits the possibility that volatility stays in the same level, although
the jump (that is, the announcement) has taken place. We have also presented plots with actual examples of these solutions, computed using Monte Carlo numerical techniques.

\section*{Acknowledgments}
\noindent
I wish to thank Josep Perell\'o for addressing me to the original Herzel's paper, and for many valuable discussions. The comments of Arturo Kohatsu-Higa and Jaume Masoliver have improved the manuscript, doing it more readable. AKH also encouraged me to find explicit analytic solutions. 
This work has been supported in part by Direcci\'on General de
Investigaci\'on under contract No. BFM2000-0795 and by Generalitat de Catalunya under contract No. 2001SGR-00061.

\renewcommand{\theequation}{\Alph{section}.\arabic{equation}}
\setcounter{equation}{0}
\appendix
\section{Appendix}
\noindent
In this appendix we present an heuristic approach trying to remove the risk associated to the volatility change in a portfolio without secondary calls. It it obvious that we will fail in this task by construction, because this is only feasible if the jump time is deterministic. But in this case, Eq.~(\ref{dC}) and the expressions derived from it are not longer correct. Therefore, we will assume that $\tau$ is a stochastic magnitude, and the solution we obtain will be conditioned to its value. Let see how it works just recovering Eq.~(\ref{dY}), and setting  $\bar{\psi}=0$:
\begin{equation*}
dY=\Delta C\left(\chi dt + d{\mathbf 1}_{t\geq \tau}\right).  
\end{equation*}
We are now demanding that $dY=0$. This constrain leads to:
\begin{equation*}
\chi=-\frac{d}{d t} {\mathbf 1}_{t\geq \tau} = -\delta(t-\tau),
\end{equation*}
and to the corresponding equation is then:
\begin{equation}
\partial_t C +\frac{1}{2} \sigma^2 S^2 \partial_{SS}^2 C -rC +rS\partial_S C +\Delta C  \delta(t-\tau)=0. \label{H_eq}
\end{equation}
Note that this equation corresponds to Eq.~(\ref{PDE_1}) with $\delta=\partial_S C$ and $\psi=0$, as we have just stated. We will search for a solution of the form 
\begin{equation*}
C(t,S,\sigma;K)=e^{-r(T-t)} U\left(T-t,\log(S)+\left(r-\frac{\sigma^2}{2}\right)(T-t);K\right),
\end{equation*}
based upon the two variable function $U(t^*,x;K)$, which must fulfil the following partial equation:
\begin{equation*}
\partial_{t^*}  U-\frac{1}{2} \sigma^2 \partial_{xx}^2 U=\Delta U \delta(T-\tau -t^*).
\end{equation*}
We therefore consider the Fourier-Laplace transform of $U(t^*,x;K)$, 
\begin{equation*}
\widehat{U}(s,\omega;K)=\int_0^{+\infty}dt^* e^{-st^*}\int_{-\infty}^{+\infty} dx \, e^{i \omega x} \, U(t^*,x;K),
\end{equation*}
that follows the simpler equation:
\begin{equation*}
s\widehat{U}-\widetilde{U}_0 +\frac{1}{2} \sigma^2 \omega^2 \widehat{U}-\Delta \widetilde{U}_{T-\tau} e^{-s(T-\tau)} {\mathbf 1}_{\tau \leq T}=0,
\end{equation*}
where the tilde stands for the Fourier transform of the corresponding object. Thus
$\widetilde{U}_0(\omega;K) \equiv \widetilde{U}(t^*=0,\omega,\sigma;K)$, and it does not depend on $\sigma$. On the other hand, $\Delta \widetilde{U}_{T-\tau}\equiv\widetilde{U}(T-\tau,\omega,\sigma_b;K)-\widetilde{U}(T-\tau,\omega,\sigma_a;K)$. Now we can isolate all the explicit dependence on the Laplace variable $s$,
\begin{equation*}
\widehat{U}(s,\omega,\sigma;K)=\frac{1}{s+\sigma^2 \omega^2/2} \left\{\widetilde{U}_0 +\Delta \widetilde{U}_{T-\tau}  e^{-s(T-\tau)} {\mathbf 1}_{\tau \leq T}\right\},
\end{equation*}
and perform an inverse transformation,
\begin{equation}
\widetilde{U}(t^*,\omega,\sigma;K)=\widetilde{U}_0 e^{-\sigma^2 \omega^2 t^*/2} +\Delta \widetilde{U}_{T-\tau} e^{-\sigma^2 \omega^2 (t^*-T+\tau)/2} {\mathbf 1}_{T-t^*< \tau \leq T}. \label{MasterU}
\end{equation}
Notoriously, the second term only gives contribution when the jump is comprised between $t$, and the maturity, $T$.
When $t^*\leq T-\tau$, {\it i.e.\/}  $\tau\leq t$ and $\sigma=\sigma_b$, Eq.~(\ref{MasterU}) reduces to,
\begin{equation*}
\widetilde{U}(t^*,\omega,\sigma_b;K)=\widetilde{U}_0 e^{-\sigma_b^2 \omega^2 t^*/2}=\widetilde{U}^{\BS}(t^*,\omega,\sigma_b;K),
\end{equation*}
which leads to
\begin{equation*}
C(t,S,\sigma_b;K)=C^{\BS}(t,S,\sigma_b;K),
\end{equation*}
a riskless price.

When $t^*>T-\tau$, that is to say, when $t<\tau$ and $\sigma=\sigma_a$, but $\tau>T$, the main equation also takes a simple form,
\begin{equation*}
\widetilde{U}(t^*,\omega,\sigma_a;K)=\widetilde{U}_0 e^{-\sigma_a^2 \omega^2 t^*/2}=\widetilde{U}^{\BS}(t^*,\omega,\sigma_a;K).
\end{equation*}
In this case, since the change in volatility occurs {\it after\/} the maturity of the contract, the price reduces to a plain Black-Scholes model without any jump of volatility,
\begin{equation*}
C(t,S,\sigma_a;K|T<\tau)=C^{\BS}(t,S,\sigma_a;K),
\end{equation*}
Thus, this scenario has again no risk associate with it.

Finally, when $t^*>T-\tau$ and $\tau \leq T$, all the terms contribute to a more complex expression,
\begin{equation}
\widetilde{U}(t^*,\omega,\sigma_a;K)=\widetilde{U}_0 e^{-\sigma_a^2 \omega^2 t^*/2} +\Delta \widetilde{U}_{T-\tau} e^{-\sigma_a^2 \omega^2 (t^*-T+\tau)/2}. \label{complexU}
\end{equation}
Recall that $\, \Delta \widetilde{U}_{T-\tau} \equiv \widetilde{U}(T-\tau,\omega,\sigma_b;K) -\widetilde{U}(T-\tau,\omega,\sigma_a;K)$
is a term that counts {\it only\/} for the variation in $\widetilde{U}$ due to the change in the volatility, when it takes place. Thus $\widetilde{U}(T-\tau,\omega,\sigma_b;K)=\widetilde{U}^{\BS}(T-\tau,\omega,\sigma_b;K)$. The other term can be obtained by self-consistency. We will start from Eq.~(\ref{complexU}) and take a limit:
\begin{eqnarray*}
\widetilde{U}(T-\tau,\omega,\sigma_a;K)&=&\lim_{t^* \rightarrow T-\tau}\widetilde{U}(t^*,\omega,\sigma_a;K)\\
&=&\lim_{t^* \rightarrow T-\tau}\widetilde{U}_0 e^{-\sigma_a^2 \omega^2 t^*/2} +\Delta \widetilde{U}_{T-\tau} e^{-\sigma_a^2 \omega^2 (t^*-T+\tau)/2},
\end{eqnarray*}
which leads to
\begin{equation*}
\widetilde{U}(T-\tau,\omega,\sigma_a;K)=\frac{1}{2}\left[\widetilde{U}^{\BS}(T-\tau,\omega,\sigma_a;K)+\widetilde{U}^{\BS}(T-\tau,\omega,\sigma_b;K) \right].
\end{equation*}
Now we will introduce this result back into Eq.~(\ref{complexU}), and obtain so
\begin{equation*}
\widetilde{U}(t^*,\omega,\sigma_a;K)=\frac{1}{2}\widetilde{U}_0 \left[e^{-\sigma_a^2 \omega^2 t^*/2} + e^{- {\bar{\sigma}}^2(t^*-T+\tau,t^*) \omega^2 t^*/2}\right], 
\end{equation*}
where $\bar{\sigma}(t_a,t_b)$ is the same function which we have previously defined in Eq.~(\ref{bar_s}). Therefore
\begin{equation*}
C(t,S,\sigma_a;K|t<\tau\leq T)=\frac{1}{2}\left[C^{\BS}(t,S,\sigma_a;K)+C^{\BS}(t,S,{\bar{\sigma}}(\tau-t,T-t);K)\right],
\end{equation*}

Finally, in order to obtain a expression for $t<\tau$ that does not depend on future information, we will compute the expected value of the previous conditioned solutions: 
\begin{eqnarray*}
C(t,S,\sigma_a;K)&=&E\left[C(t,S,\sigma_a;K|\tau=u)\right]\\
&=& \frac{\lambda}{2}\int_{t}^{T} du\left[C^{\BS}(t,S,\sigma_a;K)+C^{\BS}(t,S,{\bar{\sigma}}(u-t,T-t);K)\right]e^{-\lambda(u-t)}\\
&+&\lambda\int_{T}^{+\infty} du\, C^{\BS}(t,S,\sigma_a;K), e^{-\lambda(u-t)}
\end{eqnarray*}
an expression that reduces to
\begin{equation*}
C(t,S,\sigma_a;K)=C^{\BS}(t,S,\sigma_b;K) + \int_{t}^{T} du\, \frac{1+e^{-\lambda(u-t)}}{2} \partial_u C^{\BS}(t,S,\bar{\sigma}(u-t,T-t);K).
\end{equation*}
Then the complete result is
\begin{equation*}
C(t,S,\sigma;K)=C^{\BS}(t,S,\sigma_b;K) + {\mathbf 1}_{t < \tau} \int_{t}^{T} du\, \frac{1+e^{-\lambda(u-t)}}{2} \partial_u C^{\BS}(t,S,\bar{\sigma}(u-t,T-t);K),
\end{equation*}
which does not fulfil Eq.~(\ref{H_eq}), but it is still a valid solution.


\begin{thebibliography}{000}
\bibitem{BR96} I. Bajeux-Besnainou and J. C. Rochet, {\it Dynamic Spanning: are Options an Appropriate Instrument?}, Mathematical Finance {\bf 6} (1996) 1--16.
\bibitem{BS73} F. Black and M. Scholes, {\it The Pricing of Options and Corporate Liabilities}, J. Pol. Econ. {\bf 81} (1973) 637--659.
\bibitem{C64} P. H. Cootner (ed.), {\it The Random Character of
Stock Market Prices}, MIT Press, Cambridge MA (1964).
\bibitem{CR76} J. C. Cox and S. A. Ross, {\it The valuation of options for alternative stochastic processes}, J. Financial Economics {\bf 3} (1976) 145--166.
\bibitem{JR96} J. Jackwerth and M. Rubinstein, {\it Recovering probability distributions from contemporaneous security  prices}, J. Finance {\bf 51} (1996) 1611--1631. %
\bibitem{HP81}
M. Harrison and S. Pliska, {\it Martingales and stochastic integrals in the theory of continuos trading}, Stochastic Process. Appl. {\bf 11} (1981) 215--260.
\bibitem{H98} S. Herzel, {\it A simple model for option pricing with jumping stochastic volatility}, Int. J. Theoretical and Applied Finance {\bf 1} (1998) 487--505.
\bibitem{H93} S. L. Heston, {\it A closed-form solution for options with stochastic volatility with applications to bond and currency options}, Rev. Financial Studies {\bf 6} (1993) 327--343.
\bibitem{HW87} J. Hull and A. White, {\it The pricing of options on assets with stochastic volatilities}, J. Finance {\bf 42} (1987) 281--300.
\bibitem{MP02} J. Masoliver and J. Perell\'o, {\it A correlated stochastic volatility model measuring leverage and other stylized facts}, Int. J. Theoretical and Applied Finance {\bf 5} (2002) 541--562. 
\bibitem{MP03} J. Masoliver and J. Perell\'o, {\it Option pricing and perfect hedging on correlated stocks}, Physica A (in press). 
\bibitem{M73} R. C. Merton, {\it Theory of rational option pricing}, Bell J. Econ. and Management Sci. {\bf 4} (1973) 141--183.
\bibitem{M76} R. C. Merton, {\it Option  Pricing when underlying stock returns are discontinuous}, J. Financial Economics {\bf 3} (1976) 125--144.
\bibitem{N93} V. Naik, {\it Option Valuation and Hedging Strategies with Jumps in the Volatility of Asset Returns}, J. Finance {\bf 48} (1993) 1969--1984.
\bibitem{S87} L. O. Scott, {\it Option Pricing when the Variance Changes Randomly: Theory, Estimation and an Application}, J. financial and Quantitative Anal. {\bf 22} (1987) 419--438.
\bibitem{SS91} E. M. Stein and J. C. Stein, {\it Stock Price Distributions with Stochastic Volatility: An Analytic Approach}, Rev. Financial Studies {\bf 4} (1991) 727--752.
\bibitem{W87} J. B. Wiggins, {\it Option Values under Stochastic Volatility: Theory and Empirical Estimates}, J. Financial Economics {\bf 19} (1987) 351--372. 
\bibitem{W98} P. Wilmott, {\it Derivatives}, John Wiley \& Sons, Chichester (1998).
\end{thebibliography}
\end{document}